\documentclass[twoside]{article}
\usepackage{fleqn,espcrc2}
\usepackage{epsfig}
\usepackage{graphicx}
\usepackage[figuresright]{rotating}

\def\csw{c_{\rm sw}}

\def\fm{\rm fm}

\newcommand{\eqn}[1]{Equation~(\ref{#1})}

\newcommand{\fig}[1]{Figure~\ref{#1}}

 \title{$N^*$ Spectrum
using an ${\cal O}(a)$-Improved Fermion Action}
\author{D.G.~Richards\address{Jefferson Laboratory MS 12H2, 12000
Jefferson Avenue, \\Newport News, VA 23606, USA}\address{Old Dominion
University, Norfolk, VA 23529, USA}\thanks{This work was supported by
DOE contract DE-AC05-84ER40150 under which SURA operates the Thomas
Jefferson National Accelerator Facility}, \textit{LHPC} and
\textit{UKQCD} Collaborations}
\begin{document}
\begin{abstract}
     The construction of operators and calculational methods for the
     determination of the $N^*$ spectrum are discussed. The masses of
     the parity partners of the nucleon and delta are computed from
     the ${\cal O}(a)$-improved data of the UKQCD Collaboration, and a
     clear splitting observed between the mass of the nucleon and its
     parity partner. 
\vspace{1pc}
\end{abstract}
\maketitle
\section{INTRODUCTION}
There has been burgeoning interest in the calculation of the excited
spectrum of nucleons, spurred by the experimental programmes at
Jefferson Laboratory and elsewhere.  Coinciding with this experimental
interest, there has been a flurry of activity in the lattice
community.  Two calculations of the mass of the parity partner of the
nucleon have appeared; the first employed the highly-improved
$D_{234}$ fermion action\cite{lee98,lee00}, whilst the second employed
domain-wall fermions\cite{sasaki99}. Both calculations
exhibited a clear splitting between the masses of the $N^{1/2+}$
and $N^{1/2-}$ states, and the latter calculation stressed the
importance of chiral symmetry in obtaining a non-zero mass splitting.
In this talk, I will present a calculation of the masses of the lowest
lying negative-parity baryons obtained using the non-perturbatively
improved clover fermion action.

\section{CALCULATIONAL DETAILS}
There are three local interpolating operators employed for a
positive-parity nucleon at rest:
\begin{eqnarray}
N_1^{1/2+} & = & \epsilon_{ijk} (u_i^T C \gamma_5 d_j) u_k\label{eq:N1}\\
N_2^{1/2+} & = & \epsilon_{ijk} (u_i^T C d_j) \gamma_5 u_k\label{eq:N2}\\
N_3^{1/2+} & = & \epsilon_{ijk} (u_i^T C \gamma_4 \gamma_5 d_j) u_k.
\label{eq:N3}
\end{eqnarray}
The ``diquark'' part of both $N_1$ and $N_3$ couples upper spinor
components, whilst that in $N_2$ involves lower components and vanishes
in the non-relativistic limit.  In practice, $N_1$ and $N_3$ have a
much greater overlap with the nucleon ground state than $N_2$, and are
used in spectroscopy calculations to obtain the nucleon mass.  The
operators of eqns.~(\ref{eq:N1})-(\ref{eq:N3}) are appropriate for
positive-parity states; interpolating operators for negative-parity
states are constructed by simply multiplying by $\gamma_5$. 

Correlators constructed from these
operators receive contributions from states of both parities.
The best delineation that can be achieved is that of forward-propagating
positive-parity states and backward-propagating negative-parity
states, or the converse, through the use of the parity projection
operator $(1 \pm \gamma_4)$.  On a lattice anti-periodic in time, the
correlator may be written
\begin{eqnarray}
\lefteqn{
C_{N_i^{+/-}}(t) = \sum_{\vec{x}} \left( (1\pm\gamma_4)_{\alpha\beta}
\langle N_{i,\alpha}(\vec{x},t) \overline{N}_{i,\beta}(0) \rangle +
\right.} \nonumber\\
& & \left. (1 \mp \gamma_4)_{\alpha\beta} \langle N_{i,\alpha}(\vec{x}, N_t -
t) \overline{N}_{i,\beta}(0)\rangle \right),\label{eq:corrs}
\end{eqnarray}
where $N_t$ is the temporal extent of the lattice.
At large distances, when $t \gg 1$ and $N_t -t \gg 1$, the correlators
behave as
\begin{eqnarray}
C_{N_i^+}(t) & \rightarrow & A_i^+ e^{-M_i^+ t} + A_i^- e^{-M_i^-(N_t - t)} \\
C_{N_i^-}(t) & \rightarrow & A_i^- e^{-M_i^- t} + A_i^+ e^{-M_i^+(N_t -
t)}
\end{eqnarray}
where $M_i^+$ and $M_i^-$ are the lightest positive- and
negative-parity masses respectively.  In the presence of unbroken
chiral symmetry, the positive- and negative-parity baryons would form
mass-degenerate doublets.  The mass splitting between them is a
manifestation of spontaneously broken chiral symmetry.

A local interpolating operator for the lowest-lying spin states of the
$I=3/2$ $\Delta$ baryon is
\begin{equation}
\Delta^{3/2,1/2} = \epsilon_{ijk}(u^T_i C
\gamma_{\mu} u_j) u_k.
\end{equation}
This has an overlap onto both spin-$3/2$ and spin-$1/2$ states, but
these can be distinguished using a suitable projection, and the
positive- and negative-parity states delineated as described above.

\section{CALCULATIONAL DETAILS}
The calculation is performed using the gauge configurations and
propagators generated by the UKQCD Collaboration.  Two values of the
coupling, $\beta = 6.0$ and $\beta = 6.2$, are employed, and
propagators are computed using the clover fermion action, with the
clover coefficient determined non-perturbatively thus removing all
${\cal O}(a)$ discretisation errors. The quark propagators are
computed from both local and fuzzed sources to local and fuzzed sinks.
The parameters of the calculation are summarised in
Table~\ref{tab:params}, and further details are contained in
ref.~\cite{ukqcd99}.
\begin{table*}[htb]
\caption{The parameters of the lattices used in the calculation.}
\label{tab:params}
\begin{tabular}{cccccc}
$\beta$ & $\csw$ & $L^3\cdot T$ & $L\,[\fm]$ & $\kappa$
        & {\#conf.} \\
\hline
6.0 & 1.769 & $16^3\cdot48$ & 1.5 & $0.13344,\,0.13417,\,0.13455$ & 496\\
6.2 & 1.614 & $24^3\cdot48$ & 1.6 & $0.13460,\,0.13510,\,0.13530$ &
        216\\
\hline
\end{tabular}
\end{table*}

\begin{figure}[htb]
\hspace{-0.4cm}\epsfxsize=220pt\epsfbox{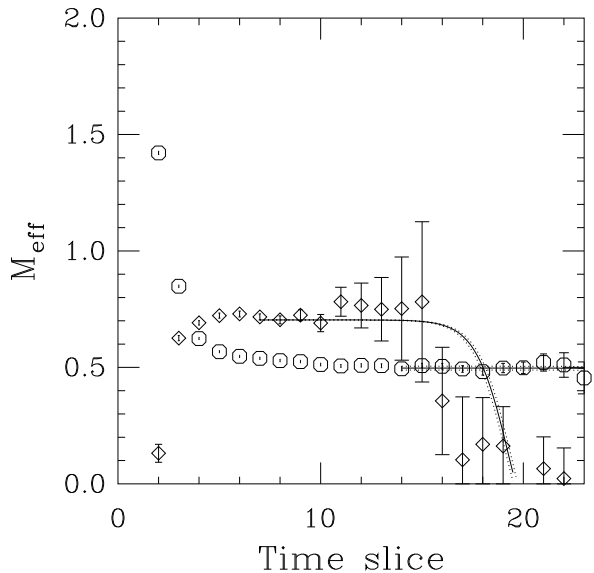}
\caption{The effective masses for the $N^{1/2+}$ channel (circles) and
the $N^{1/2-}$ channel (diamonds) at $\beta = 6.2$
with $\kappa= 0.1351$.  The lines are from a simultaneous fit to both
parities.}
\label{fig:effmass}
\end{figure}
The masses of the $N^{1/2+}$ and $N^{1/2-}$ states are obtained from a
simultaneous, four-parameter fit to the correlators $C_{N_1^{+}}(t)$
and $C_{N_1^{-}}(t)$ of \eqn{eq:corrs}, using fuzzed sources and local
sinks.  The quality of the data and of the fits for both the positive-
and negative-parity states is illustrated in \fig{fig:effmass}, and
the importance of including the backward-propagating positive-parity
state in the determination of the negative-parity mass is clear.

\begin{figure}[htb]
\hspace{-0.4cm}\epsfxsize=220pt\epsfbox{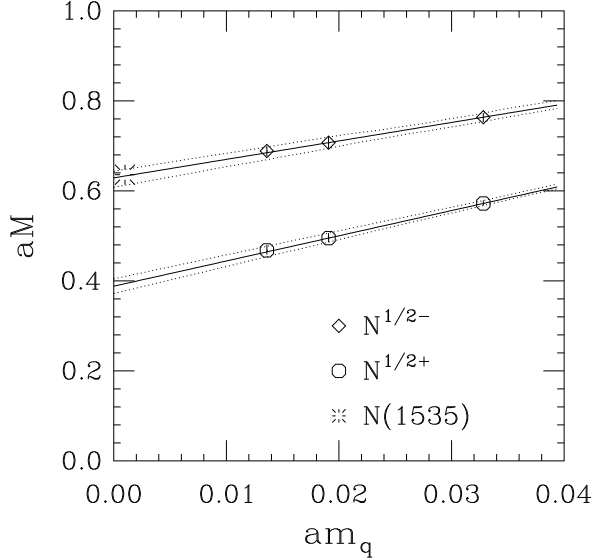}
\caption{The masses of the $N^{1/2+}$ (circles) and $N^{1/2-}$
(diamonds) at $\beta = 6.2$.  The lines are linear extrapolations to
the chiral limit.  The burst is the physical $N(1535)$ mass, expressed
in lattice units as discussed in the text.}\label{fig:chiral_extrap}
\end{figure}

The masses in the chiral limit are obtained by linear extrapolation in
the quark mass, as shown in \fig{fig:chiral_extrap} for the data at
$\beta = 6.2$.  The results show a clear splitting between the masses
of the positive- and negative-parity states persisting to the chiral
limit. Also shown as the burst is the experimental determination of
the lowest lying $J = 1/2^{-}$ mass, $N(1535)$, scaled with the
lattice spacing determined from the nucleon mass; the agreement
between the lattice determination of the $N^{1/2-}$ mass and the
experimental value is striking.  The quality of the extracted masses
is poorer at $\beta = 6.0$ than at $\beta = 6.2$, reflecting the
larger value of the masses in lattice units.  As observed in
ref.~\cite{sasaki99}, the ``bad'' baryon operator $N_2$ yields a much
higher mass for the positive-parity ground state, comparable with that
of the negative-parity baryon obtained using $N_1$.

\begin{figure}[htb]
\hspace{-0.4cm}\epsfxsize=220pt\epsfbox{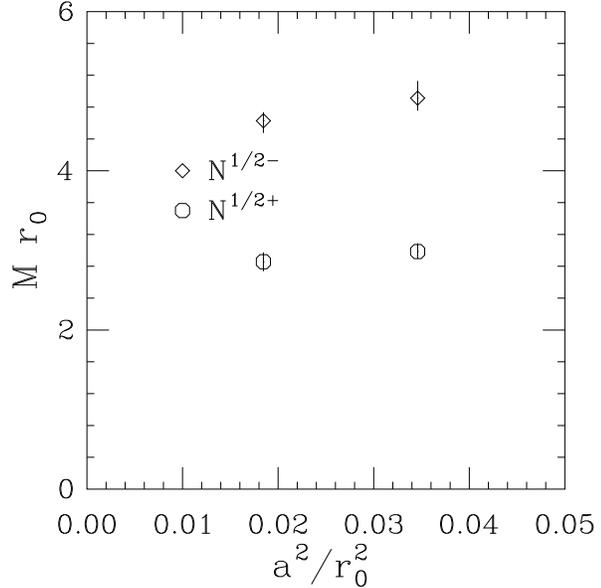}
\caption{The masses of the $N^{1/2+}$ and $N^{1/2-}$ states in units
of $r_0$ against the lattice spacing $a^2$ in units of $r_0^2$.}
\label{fig:continuum}
\end{figure}
With only two values of the lattice spacing, it is not possible to
perform a continuum extrapolation.  However some indication of the
magnitude of discretisation effects can be gleaned from
\fig{fig:continuum}, where the masses of the $N^{1/2+}$ and $N^{1/2-}$
states are shown in units of Sommer's $r_0$ at each of the two lattice
spacings. The higher spin states  are expected to
be larger, and thus more susceptible to finite-volume corrections. A
analysis including data at three values of the lattice spacing and at
different lattice volumes is in
progress\cite{qcdsf00}.

\begin{figure}[htb]
\hspace{-0.4cm}\epsfxsize=220pt\epsfbox{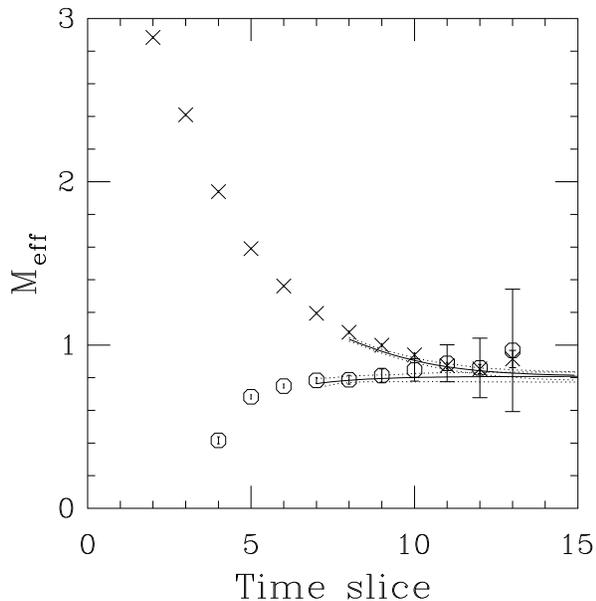}
\caption{The effective mass in the $\Delta^{3/2-}$ channel at $\beta =
6.2$ and $\kappa=0.1351$ for the local-local (crosses) and
fuzzed-local (circles) correlators.  The curves are from a simultaneous,
single-mass fit to both correlators}
\label{fig:delta_effmass}
\end{figure}
\begin{figure}[htb]
\hspace{-0.4cm}\epsfxsize=220pt\epsfbox{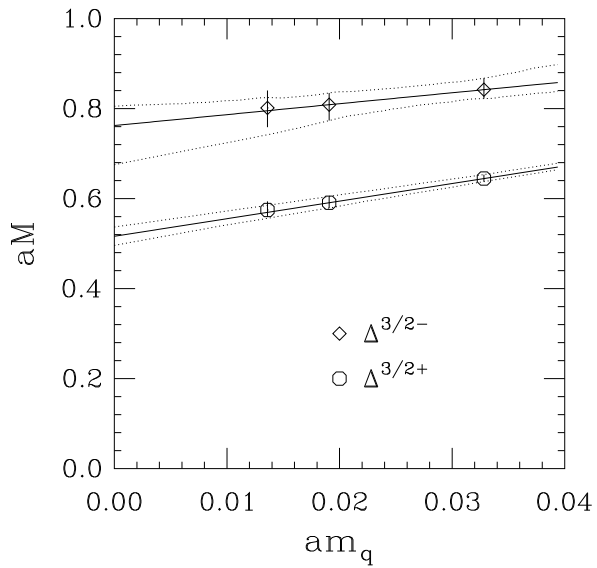}
\caption{The masses of the $\Delta^{3/2+}$ (circles) and
$\Delta^{3/2-}$ (diamonds) at $\beta = 6.2$.
The lines are linear extrapolations to
the chiral limit.}\label{fig:delta}
\end{figure}

The mass of the $\Delta^{3/2-}$ is also accessible,
though subject to greater statistical noise.  Here the
mass is obtained from a simultaneous single-mass fit to the
fuzzed-local and local-local correlators in which the contribution of
the opposite-parity state is neglected, as illustrated in
\fig{fig:delta_effmass}.  A clear splitting is seen between the masses
of the positive- and negative-parity states, shown in \fig{fig:delta}.

\section{DISCUSSION}
This calculation using the non-perturbatively improved
clover fermion action is capable of resolving the mass splitting
between positive- and negative-parity baryon states, and yields a
value consistent with experiment.  Under the
$SU(6)\otimes O(3)$ symmetry of spin-flavour and orbital angular
momentum, the low-lying negative parity baryons are assigned to the
$\underline{70}$-plet representation.  Calculations of the masses of
these states both in the quark model\cite{isgur_karl}, and in large
$N_C$\cite{carlson99}, suggest that the spin-orbit contribution is
surprisingly small, whilst the hyperfine contribution is crucial.  The
introduction of the clover term to the Wilson fermion action improves
the prediction for hyperfine splittings (see
e.g. ref.~\cite{ukqcd93}), and this may account for earlier failures
to present measurements of the negative-parity masses using the Wilson
fermion action.  A comparison of results using the two actions will be
included in a later paper\cite{qcdsf00}.

The determination of the masses of spin states above $3/2$ requires
the use of non-local baryon operators.  Indeed, baryon spins above
$5/2$ are contaminated by lower spins in the same irreducible
representations of the cubic group\cite{johnson82}. Nevertheless, the
\textit{ab initio} determination of the masses of the first few
spin/parities of the $N^*$ system seems feasible, with the promise of
invaluable information about the nature of QCD and hadronic physics.

\section*{ACKNOWLEDGEMENTS}
I am very grateful for invaluable discussions with Robert Edwards, Jose Goity,
Nathan Isgur, Richard Lebed, Frank Lee, and Stephen Wallace.

\end{document}